\documentclass[sigconf,screen, nonacm]{acmart}
\AtBeginDocument{%
  }

\acmYear{2026}
\copyrightyear{2026}
\setcopyright{acmlicensed}
\acmDOI{XXXXXXX.XXXXXXX}

\acmConference[Conference acronym 'XX]{Make sure to enter the correct
  conference title from your rights confirmation email}{June 03--05,
  2018}{Woodstock, NY}

\acmISBN{978-1-4503-XXXX-X/2018/06}

\usepackage{tabularx}
\usepackage{tcolorbox}
\usepackage{caption}
\usepackage{booktabs}
\usepackage{multirow}
\usepackage{makecell}
\tcbset{
  mybox/.style={
    colframe=gray!70,
    colback=white,
    colbacktitle=gray!15,  
    coltitle=black,        
    fonttitle=\bfseries\small,
    boxrule=0.5pt,
    arc=2pt,
    left=3pt,right=3pt,top=3pt,bottom=3pt,
    before skip=3pt, after skip=3pt,
    title filled          
  }
}
\begin{document}

\title{A Differential Fuzzing-Based Evaluation of Functional Equivalence in LLM-Generated Code Refactorings}

\author{Simantika Bhattacharjee Dristi}

\affiliation{%
  \institution{University of Virginia}
  \city {Charlottesville}
  \country{USA}}
\email{nwc8gr@virginia.edu}
\orcid{0009-0007-5511-9307}

\author{Matthew B. Dwyer}
\affiliation{%
  \institution{University of Virginia}
  \city {Charlottesville}
  \country{USA}}
\email{matthewbdwyer@virginia.edu}
\orcid{0000-0002-1937-1544}

\begin{abstract}
With the rapid adoption of large language models (LLMs) in automated code refactoring, assessing and ensuring functional equivalence between LLM-generated refactoring and the original implementation becomes critical. While prior work typically relies on predefined test cases to evaluate correctness, in this work, we leverage differential fuzzing to check functional equivalence in LLM-generated code refactorings. Unlike test-based evaluation, a differential fuzzing-based equivalence checker needs no predefined test cases and can explore a much larger input space by executing and comparing thousands of automatically generated test inputs.  In a large-scale evaluation of six LLMs (CodeLlama, Codestral, StarChat2, Qwen-2.5, Olmo-3, and GPT-4o) across three datasets and two refactoring types, we find that LLMs show a non-trivial tendency to alter program semantics, producing \textbf{19-35\%} functionally non-equivalent refactorings. Our experiments further demonstrate that about \textbf{21\%} of these non-equivalent refactorings remain undetected by the existing test suites of the three evaluated datasets. Collectively, the findings of this study imply that reliance on existing tests might overestimate functional equivalence in LLM-generated code refactorings, which remain prone to semantic divergence.
\end{abstract}





\maketitle

\section{Introduction}
\begin{figure*}[t]
  \centering
  \includegraphics[width=0.9\textwidth]{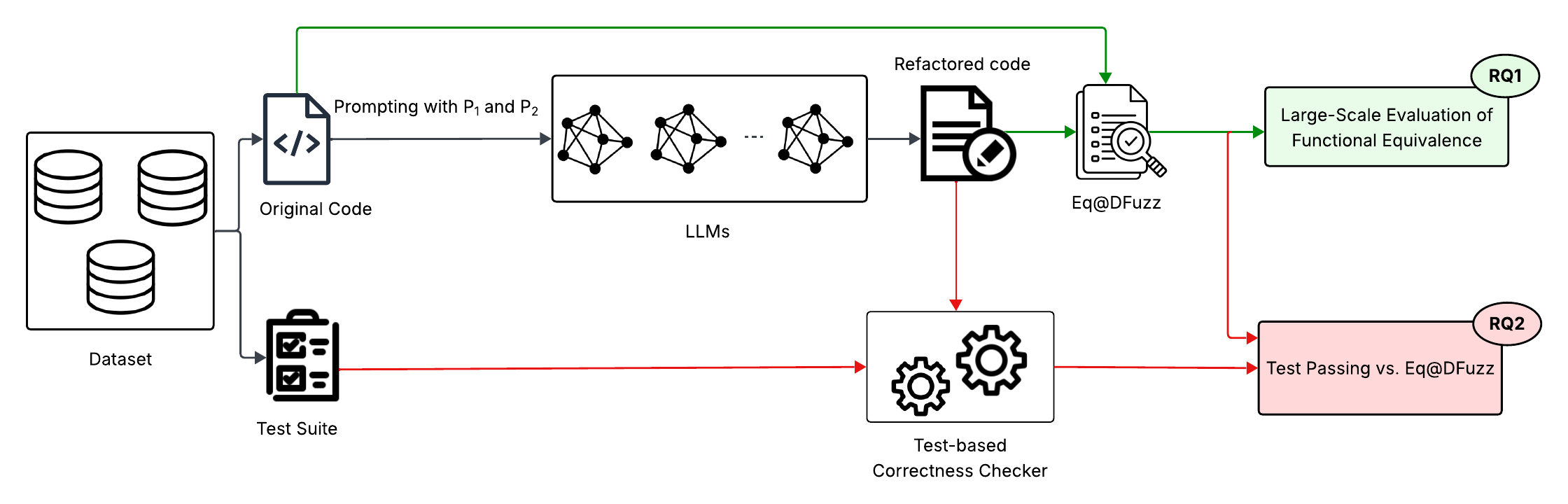}
  \vspace{-5mm}
  \caption{Overview of our approach}
  \label{fig:approach}
\end{figure*}
Code refactoring refers to the task of restructuring code to improve its internal quality while preserving its functional behavior ~\cite{fowler_book}. Automated code refactoring is gaining traction as it improves the quality of existing code without requiring manual rewrites, thus reducing both development effort and the risk of human error ~\cite{LLM_refactorings, pyrefact, rewrite_python, PIE, refactoring_few_shot}.  While automated refactoring has long relied on rule-based tools ~\cite{pyrefact, rewrite_python}, recent years have seen increasing adoption of large language models (LLMs) due to their ability to perform a broader range of code rewrites beyond specific transformation rules ~\cite{LLM_refactorings, PIE, supercoder, refactoring_few_shot}. This has led to the development and widespread use of several agentic frameworks for code refactoring ~\cite{refactorGPT, refAgent}. As such frameworks are increasingly integrated into the CI/CD pipeline, reliable evaluation of large language models’ refactoring capabilities becomes more important than ever. 
\par A key criterion for evaluating refactorings is correctness, i.e., the refactored code should remain functionally equivalent to the original. Thus, trusting LLMs for code refactoring largely depends on investigating \textit{how often LLM-generated refactorings alter program semantics.} However, prior work evaluating LLMs for code refactoring has largely emphasized performance-oriented metrics such as speedup, cyclomatic complexity, and lines of code ~\cite{refactoring_few_shot, cordeiro2024empiricalstudycoderefactoring}, leaving functional equivalence insufficiently examined. Although some earlier studies ~\cite{swerefactor, refactoring_few_shot, supercoder} evaluate the correctness of LLM-generated refactorings, all of them rely on test passing-based metrics such as pass@k ~\cite{chen2021evaluatinglargelanguagemodels} or test pass ratio ~\cite{APPS} as proxies for functional equivalence. This is problematic because such evaluations depend heavily on the quality of the test suite, which often covers only a limited portion of the input space ~\cite{humanevalplus}, overlooking deeper program semantics that may be altered by refactoring. As a result, a refactored code may pass all tests in the existing test suite, while still failing to be functionally equivalent to the original code. Moreover, predefined test suites may not always be available.
\par To address these gaps, in this work, we empirically evaluate the functional equivalence of code refactorings generated by six widely used LLMs -  CodeLlama \footnote{\url{https://huggingface.co/codellama/CodeLlama-13b-hf}}, Codestral \footnote{\url{https://huggingface.co/mistralai/Codestral-22B-v0.1}}, StarChat2 \footnote{\url{https://huggingface.co/HuggingFaceH4/starchat-beta}}, Qwen-2.5 \footnote{\url{https://huggingface.co/Qwen/Qwen2.5-Coder-32B}}, Olmo-3 \footnote{\url{https://huggingface.co/allenai/Olmo-3-1125-32B}}, and GPT-4o \footnote{\url{https://openai.com/index/hello-gpt-4o/}} - across three datasets (HumanEval ~\cite{chen2021evaluatinglargelanguagemodels}, MBPP ~\cite{MBPP}, and APPS ~\cite{APPS}) and two refactoring types: performance optimization and code simplification. Instead of relying on a predefined test suite, we adopt the differential fuzzing approach introduced in ~\cite{LoCaL} to define an equivalence-checking method, Eq@DFuzz. In this method, a fuzzer generates a diverse set of test inputs, and each input is executed on both the original and refactored code. The refactored code is considered functionally equivalent to the original only if their outputs match across all generated inputs. Eq@DFuzz eliminates reliance on predefined test suites and provides a more reliable estimate of equivalence by generating a much larger number of test inputs that significantly expand the explored input space. Experimental results in Section \ref{sec:RQ1} demonstrate that \textbf{all of the evaluated LLMs generate a substantial fraction (19-35\%) of non-equivalent refactorings}, raising concerns regarding the trustworthiness of LLM-generated code refactorings. Our findings also indicate that the likelihood of generating non-equivalent refactorings increases with the complexity of the code to be refactored. On the more complex dataset APPS, the proportion of refactorings that break existing functionality is 7 and 10 percentage points higher than on MBPP and HumanEval, respectively.
\par We further evaluate our hypothesis that reliance on test passing overestimates the functional equivalence of LLM-generated refactorings. To this end, we compare test-suite-based correctness, with equivalence measured via Eq@DFuzz. As the empirical evidence in Section \ref{sec:RQ2} shows, \textbf{test passing is an unreliable proxy for functional equivalence, as it fails to detect a significant fraction ($\mathbf{\approx 21\%}$) of non-equivalent refactorings.} Thus, the findings of this paper emphasize the need to move beyond predefined test suites and adopt stronger equivalence-checking techniques, such as differential fuzzing, to reliably assess the correctness of LLM-generated refactorings.

\section{Approach}

Fig. \ref{fig:approach} shows a general overview of our approach. In this section, we first discuss the models, datasets, and prompts that are used in this study, and then detail the differential fuzzing-based equivalence checking method, Eq@DFuzz. 
\subsection{Models, Datasets and Prompts} 
\subsubsection{Large Language Models} We analyze the functional equivalence of refactorings generated by six LLMs - five open-source models (CodeLlama, Codestral, StarChat2, Olmo-3, and Qwen-2.5) and the closed-source GPT-4o. 
The open-source models are all decoder-only transformers trained on large-scale multilingual code corpora, differing in their underlying architectural families and ranging from 13B to 32B parameters in size.\\
\textbf{CodeLlama:} We utilize the 13B parameter variant of CodeLlama, a decoder-only language model from the LLaMA family trained for code generation and comprehension. We include CodeLlama due to its strong performance on code generation and code infilling tasks ~\cite{rozière2024codellamaopenfoundation} and its widespread use as a baseline open-source code LLM.\\
\textbf{Codestral:} In this study, we evaluate the publicly available 22B Codestral model from Hugging Face ~\cite{huggingface}. Codestral is selected to represent the Mistral model family, which differs architecturally from LLaMA-based models.\\
\textbf{StarChat2:} StarChat is a series of GPT-like language models fine-tuned to support both chat and programming tasks. We utilize the 15B version of StarChat2, the latest model in the StarChat series. Prior work ~\cite{cordeiro2024empiricalstudycoderefactoring} has demonstrated its strong performance on code refactoring, motivating its inclusion in our analysis. \\
\textbf{Qwen-2.5-Coder:} The Qwen-2.5-Coder family comprises six variants of sizes between 0.5B and 32B, specializing in code-centric tasks like code generation, code reasoning, and code fixing. The 32B variant is widely regarded as a strong open-source model, with reported coding performance comparable to GPT-4o ~\cite{qwen25coder32binstruct}. Using this model in our study enables an evaluation of semantic robustness in refactorings generated by state-of-the-art open-source codeLLMs.\\
\textbf{Olmo-3:} Developed by the Allen Institute for AI (AI2) ~\cite{ai2docs}, Olmo-3 is a newly introduced model whose weights, training data, and training code are publicly released to promote transparency. In this work, we use the 32B variant ~\cite{olmo3}, which is reported to perform close to state-of-the-art on coding benchmarks such as HumanEval ~\cite{chen2021evaluatinglargelanguagemodels} and BigCodeBench ~\cite{bigcodebench}. As a recent release, Olmo3 remains relatively under-audited, making it a relevant model for evaluation in our study.\\
\textbf{GPT-4o:} GPT-4o is a closed-source, multimodal, and multilingual LLM developed by OpenAI, with strong reported performance on the widely used benchmark, HumanEval ~\cite{gpt_4o}. We access GPT-4o through the OpenAI API in our experiments.
\subsubsection{Datasets} 
\begin{table}[t]
\centering
\caption{Datasets used in our evaluation. ``Total'' refers to all problems in the dataset, and ``k'' refers to the problems analyzed.}
\label{tab:datasets}
\vspace{-2mm}
\renewcommand{\arraystretch}{1.05}
\small
\begin{tabularx}{\columnwidth}{l c c c c}
\toprule
\textbf{Dataset} & \textbf{\# Total} & \textbf{k} & \textbf{\# Test Cases/Problem} & \textbf{Level} \\
\midrule
HumanEval & 164  & 164 & 7.7  & Function \\
MBPP      & 974  & 100 & 3.0  & Function \\
APPS      & 5000 & 100 & 21.2 & Program  \\
\bottomrule
\end{tabularx}
\vspace{-4mm}
\end{table}

To measure functional equivalence in code refactorings, we use the test splits of three popular coding benchmarks: HumanEval and MBPP with function-level codes and APPS with program-level codes. The inclusion of both function-level and program-level code allows an analysis of whether code complexity influences the likelihood of functional divergence during refactoring. Each dataset includes a reference implementation that serves as the original code to be refactored in our experiments. All three datasets also provide fixed test suites, which allow for a comparison between test passing and differential-fuzzing-based equivalence to assess whether existing tests are sufficient to detect semantic breakage. Considering the runtime overhead arising from the large number of test executions required for differential fuzzing, we sample 100 entries from APPS and MBPP while using all 164 entries of HumanEval. 
Table \ref{tab:datasets} details the datasets included in our analysis.
\subsubsection{Prompts} 
While code refactoring spans a wide range of code-quality improvements, including refactoring for readability, maintainability, simplification, modernization, or performance optimization ~\cite{refactoring_few_shot, cordeiro2024empiricalstudycoderefactoring}, this work focuses on two types of refactoring: code simplification and performance optimization. Together, these types reflect common refactoring goals and capture distinct rewrite patterns - simplification typically involves less complex surface-level edits, whereas optimization often requires deeper structural transformations.

We, therefore, design two instruction-based prompts (Fig. \ref{fig:prompts}), a prompt for performance optimization, $P_1$, and a prompt for code simplification, $P_2$. We zero-shot prompt all six LLMs using both $P_1$ and $P_2$ to generate refactorings for each selected entry across three datasets, yielding a total of 4368 refactorings. Decoding parameters and the random seed are held fixed across all runs. The functional equivalence of each refactoring is then measured using a differential fuzzing–based equivalence checker (detailed in Section \ref{Diff_fuzz}).

\begin{center}
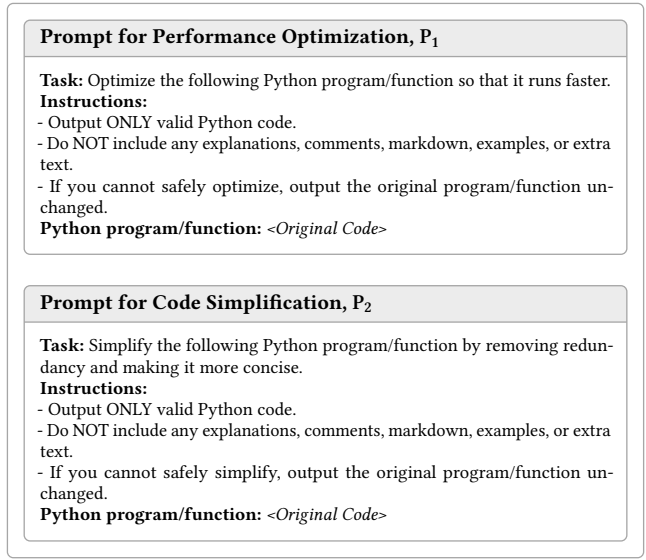


\begin{tcolorbox}[mybox]

\begin{tcolorbox}[mybox, title={Prompt for Performance Optimization, $\mathbf{P_1}$}]
\footnotesize
\textbf{Task:} Optimize the following Python program/function so that it runs faster.\\
\textbf{Instructions:}\\
- Output ONLY valid Python code.\\
- Do NOT include any explanations, comments, markdown, examples, or extra text.\\
- If you cannot safely optimize, output the original program/function unchanged.\\
\textbf{Python program/function:} \textit{<Original Code>}
\end{tcolorbox}

\vspace{0.6em}

\begin{tcolorbox}[mybox, title={Prompt for Code Simplification, $\mathbf{P_2}$}]
\footnotesize
\textbf{Task:} Simplify the following Python program/function by removing redundancy and making it more concise.\\
\textbf{Instructions:}\\
- Output ONLY valid Python code.\\
- Do NOT include any explanations, comments, markdown, examples, or extra text.\\
- If you cannot safely simplify, output the original program/function unchanged.\\
\textbf{Python program/function:} \textit{<Original Code>}
\end{tcolorbox}
\end{tcolorbox}
\vspace{-3mm}
\captionof{figure}{Prompts used to generate code refactorings with LLMs. ``Original Code" refers to the code to be refactored.}
\label{fig:prompts}
\end{center}
\subsection{Differential Fuzzing-Based Equivalence Checker, Eq@DFuzz}
\label{Diff_fuzz}
We build our equivalence checker, Eq@DFuzz, on prior work by Dristi et al. ~\cite{LoCaL}, which proposes a benchmark for evaluating code evaluation metrics using differential fuzzing–based functional similarity scores. In that approach, input constraints, such as input types, inter-dependencies, and relational conditions among arguments, are first inferred using GPT-4 and then manually validated. Atheris \cite{atheris}, a byte-level fuzzer, is used to generate a large set of test inputs satisfying these constraints. Each input is executed on both the reference implementation and its rewritten variant to compute a functional similarity score based on the proportion of output match {~\cite[Alg. 1]{LoCaL}}. Unlike ~\cite{LoCaL}, which reports continuous similarity scores, we adopt a binary (0/1) definition of equivalence, where a refactoring is considered non-equivalent as soon as any single output mismatch is detected. Following ~\cite{LoCaL}, we generate 1000 test inputs for program-level code and 2000 for function-level code and replicate all other experimental settings. Since we use the same datasets as ~\cite{LoCaL}, we reuse their inferred input constraints, eliminating the need for both constraint inference and the subsequent manual validation. 
\begin{definition}[Eq@DFuzz]
Given an original code $C$ and its refactored variant $C'$, and a set of test inputs $\mathcal{I} = \{i_1, i_2, \dots, i_n\}$ generated via fuzzing, we define Eq@DFuzz as:
\begin{equation}
\label{eq:eq-difffuzz}
\text{Eq@DFuzz}(C, C') =
\begin{cases}
1, & \text{if } \forall i \in \mathcal{I},\; C(i) = C'(i), \\
0, & \text{otherwise}.
\end{cases}
\end{equation}
Under Eq@DFuzz, a refactored program is considered functionally equivalent to the
original if and only if it produces identical outputs for all
fuzz-generated inputs; otherwise, it is classified as non-equivalent.
\end{definition}
\section{Experimental Study} The following \textit{two} research questions guide our experimental study:
\noindent\textbf{RQ1:} To what extent are LLM-generated code refactorings functionally equivalent to the original code?\\
\noindent\textbf{RQ2:} Can passing existing test suites be considered a reliable proxy for functional equivalence?
\subsection{Baseline Metric}
In Section \ref{sec:RQ2}, we compare Eq@DFuzz with traditional test-suite-based correctness. For a refactored code, $C'$, and a test suite, $\mathcal{T}$, we define correctness under the predefined test suite as, 
\begin{equation}
\text{Corr@Test} =
\mathbb{I}\left[
\forall\, t \in \mathcal{T},\; C'(t) \text{ passes}
\right],
\label{eq:corr}
\end{equation}
Corr@Test evaluates to $1$ if the refactored program $C'$ passes all tests in $\mathcal{T}$, and $0$ otherwise.

\subsection{RQ1: Assessing Functional Equivalence in LLM-generated Refactorings}
\label{sec:RQ1}
In this experiment, we conduct a large-scale comparative analysis of the refactorings generated by six LLMs across three datasets and two refactoring types. We run all 4,368 refactorings through Eq@DFuzz and remove those that result in timeouts or compilation errors, leaving 3,538 refactorings for the final analysis.
\par As shown in Table \ref{tab:rq1_results}, all six LLMs struggle to preserve program semantics during refactoring, even though they were explicitly instructed to avoid unsafe rewriting. Codestral shows the highest percentage of non-equivalent refactorings ($\approx 35\%$), followed closely by StarChat2 and CodeLlama. More concerning is that even Qwen-2.5, Olmo-3, and GPT-4o, often regarded as state-of-the-art codeLLMs, produce a non-negligible fraction of non-equivalent refactorings (19–22\%), limiting their reliability for use in automated code refactoring. We further analyze the percentage of non-equivalent refactorings across datasets and refactoring types. Among datasets, APPS has the highest percentage (32.09\%) of non-equivalent refactorings, compared to MBPP (25.33\%) and HumanEval (22.10\%). This is expected, since APPS has program-level codes, which are inherently more complex than the function-level codes in MBPP and HumanEval, making it more challenging for the LLMs to preserve program semantics while rewriting the code. Interestingly, although code simplification is considered a relatively easier refactoring that does not require as extensive structural transformations as performance optimization, the LLMs studied produce nearly the same fraction ($\approx 26\%$) of non-equivalent refactorings for both refactoring types. This raises broader concerns about the reliability of LLM-generated refactorings, given that even seemingly less complex refactorings can introduce substantial risk of semantic deviation.

\begin{table}[h]
\centering
\caption{Percentage of non-equivalent refactorings detected by Eq@DFuzz across models, datasets, and refactoring types.}
\label{tab:rq1_results}
\vspace{-2mm}
\renewcommand{\arraystretch}{1.05}
\small
\begin{tabular}{l l c c c c}
\toprule
\textbf{Model} & \textbf{Refactoring} & \textbf{HE} & \textbf{MBPP} & \textbf{APPS} & \textbf{Overall} \\
\midrule

\multirow{2}{*}{CodeLlama}
& Simplification & 33.33\%  & 24.24\%  & 26.55\% & \multirow{2}{*}{26.23\%} \\
& Optimization  & 30.95\% & 23.19\% & 15.93\% & \\
\midrule
\multirow{2}{*}{Codestral}
& Simplification & 23.81\%  & 36.07\%  & 40.35\% & \multirow{2}{*}{35.14\%} \\
& Optimization  & 27.12\% & 50.85\% & 42.11\% & \\

\midrule
\multirow{2}{*}{StarChat2}
& Simplification & 26.23\%  & 33.33\%  & 45.54\% & \multirow{2}{*}{34.24\%} \\
& Optimization  & 32.28\% & 32.20\% & 35.40\% & \\

\midrule
\multirow{2}{*}{Qwen-2.5}
& Simplification & 13.18\%  & 27.14\%  & 30.09\% & \multirow{2}{*}{22.01\%} \\
& Optimization  & 18.32\% & 18.18\% & 27.52\% & \\
\midrule
\multirow{2}{*}{Olmo-3}
& Simplification & 18.55\%  & 12.70\%  & 43.88\% & \multirow{2}{*}{21.73\%} \\
& Optimization  & 14.40\% & 8.96\% & 28.09\% & \\

\midrule
\multirow{2}{*}{GPT-4o}
& Simplification & 8.53\%  & 15.71\%  & 20.18\%  & \multirow{2}{*}{18.58\%} \\
& Optimization  & 19.69\% & 27.42\% & 28.57\% & \\

\bottomrule
\end{tabular}
\vspace{-4mm}
\end{table}

\subsection{RQ2: Test Passing vs Eq@DFuzz}
\label{sec:RQ2}
This research question investigates whether existing test suites are sufficient to identify functional non-equivalence in LLM-generated refactorings. To address this question, we compute the baseline metric, Corr@Test (Eq. ~\ref{eq:corr}) for all non-equivalent refactorings identified in Section ~\ref{sec:RQ1} and count cases where Corr@Test equals 1 while Eq@DFuzz equals 0 (Table \ref{tab:rq2_results}). For each dataset under study, the counts shown in the fourth column of the table report how often all test cases of the specific dataset pass despite the refactoring being functionally non-equivalent under differential fuzzing.  
\begin{table}[t]
\centering
\caption{Correctness under test passing (Corr@Test) vs. Equivalence under differential fuzzing (Eq@DFuzz). Total denotes all refactorings analyzed per dataset and non-eq. denotes the refactorings identified as non-equivalent by Eq@DFuzz.}
\label{tab:rq2_results}

\renewcommand{\arraystretch}{1.05}
\setlength{\tabcolsep}{6pt}
\small
\begin{tabular}{l|c|c|>{\centering\arraybackslash}p{3.2cm}}
\toprule
\textbf{Dataset} 
& \textbf{\# Total}
& \textbf{\# Non-eq.} 
& \makecell{\textbf{\# Non-eq.}\\\textbf{where Corr@Test$=\mathbf{1}$}} \\
\midrule
HumanEval    & 1507 & 333 & 69 (20.72\%) \\
\midrule
MBPP  & 766  & 194 & 42 (21.65\%) \\
\midrule
APPS  & 1265 & 406 & 91 (22.41\%) \\
\bottomrule
\end{tabular}
\vspace{-4mm}
\end{table}

\par Table \ref{tab:rq2_results} demonstrates that, on average, \textbf{21.65\%} of non-equivalent refactorings go undetected across all three datasets when correctness is assessed using existing test cases. To note, although APPS provides far more test cases per problem than MBPP and HumanEval (Table \ref{tab:datasets}), it fails to detect non-equivalence at a similar, and in fact, slightly higher rate. This suggests that the additional test cases of APPS might be redundant, tending to validate similar execution paths instead of revealing critical semantic differences. 
\par The findings of this research question have two important implications. First, relying solely on existing test cases may have led prior studies to inflated estimates of refactoring correctness. Second, the datasets analyzed in this work are widely used for evaluating LLM performance on tasks like code generation, completion, and translation ~\cite{code_comp, code_gen, code_trans}, and hence, the inadequacy of test cases identified here impacts LLM evaluation across a wide range of downstream tasks beyond refactoring alone. 
\section{Threats to Validity}
In our large-scale study of functional equivalence in LLM-generated code refactorings, we evaluate six LLMs. Although we prioritize recent SOTA models, our findings might not generalize fully to other LLMs trained with alternative architectures or strategies. While adopting prior work for differential fuzzing, we mimic their original experimental settings as much as possible to mitigate risks of implementation bugs. Reproducibility remains a challenge due to the inherent non-determinism of LLMs. However, we minimize this threat by using fixed seeds and consistent decoding parameters across all models. Our evaluation is limited to Python datasets, which may limit generalizability to other programming languages. Nonetheless, given the extensive Python-centric pretraining of modern LLMs, this choice should not unfairly disadvantage model performance. Although these datasets may not fully capture the complexity of real-world codebases, they remain widely used for evaluating LLMs in software engineering tasks. Our study demonstrates that semantic preservation is challenging even when we restrict LLM refactorings to simple examples, and the increased failure rates on APPS further indicate that the issue intensifies with scale and complexity. This suggests that our findings may underestimate the severity of the issue in large-scale code bases, though further study is needed to substantiate this claim. The findings in Section \ref{sec:RQ2} reveal a critical inadequacy of existing test-based correctness evaluation; however, we acknowledge that these conclusions could be affected by the availability of datasets with stronger test suites.
\section{Future Work}
In future work, we plan to expand our evaluation to incorporate alternative prompting strategies, such as few-shot prompting and chain-of-thought (CoT), to assess whether prompting style influences the frequency of non-equivalent refactorings. Moreover, a qualitative analysis of non-equivalent cases that pass existing test suites may help identify critical gaps and guide the design of more rigorous tests. We also aim to fine-tune open-source LLMs for reliable refactoring by simultaneously optimizing for functional equivalence and quality metrics like speedup and lines of code, so that improvements in quality do not come at the cost of correctness.
\section{Conclusion}
In this work, we present the first comprehensive study of functional equivalence in LLM-generated code refactorings. By shifting from traditional test-based validation to the differential fuzzing-based equivalence checker, Eq@DFuzz, we demonstrate that even SOTA codeLLMs break existing functionality during refactoring. Our findings also reveal that many of these non-equivalent refactorings successfully pass the test suites of three widely used coding benchmarks. Such inadequacy undermines the assumption that passing tests reliably implies functional correctness in downstream software engineering tasks like code generation, refactoring, and completion. Collectively, these results provide strong evidence that we cannot rely solely on LLMs to correctly refactor code, nor can we rely exclusively on given test suites to assess correctness.
\bibliographystyle{ACM-Reference-Format}
\bibliography{ref}

\end{document}